\documentclass[8pt]{article}

\usepackage{epsfig}
\usepackage{amsmath}
\usepackage{helvet}
\usepackage{amssymb}
\usepackage{euscript}
\usepackage{eufrak}

\newcommand{\beq}{\begin{equation}}
\newcommand{\eeq}{\end{equation}}

\newcommand{\vecE}{{\bf E}}

\newcommand{\Ham}{{\mathcal{H}}}

\newcommand{\I}{{\mathcal I}}

\newcommand{\F}{{\boldsymbol{\rm F}}}
\newcommand{\D}{{\boldsymbol{\textsc{D}}}}
\newcommand{\T}{{\textsc{T}}}
\newcommand{\n}{{\frak{n}}}
\newcommand{\nn}{{\frak{n}+1}}
\newcommand{\K}{{\mathcal{K}}}

\newcommand{\A}{{\mathcal A}}
\newcommand{\rmd}{{\rm d}}
\newcommand{\rme}{{\rm e}}
\newcommand{\e}{{\bf e}}
\newcommand{\B}{{\mathcal B}}
\begin{document}

{\Large \textbf{ \centerline{Collisionless energy absorption in nanoplasma layer}}} {\Large \textbf{\centerline{ in regular and stochastic
regimes.}}} \vskip 0.5cm \centerline{Ph~A~Korneev\footnote[0]{e-mail: korneev@theor.mephi.ru} } \vskip0.3cm \centerline{ Department of
Theoretical Nuclear Physics}, \centerline{ Moscow Engineering Physics Institute, Moscow, 115409, Russia}

\vskip1cm

\begin{abstract}
Collisionless energy absorption in 1D nanoplasma layer is considered. Straightforward classical calculation of the absorption rate in
action-angle variables is presented. In regular regime the result obtained is the same as in \cite{iop}, but deeper insight is possible now due
to the technique used. Chirikov criterion of the chaotic absorption regime is written out. Collisionless energy absorption rate in nanoplasma
layer is calculated in stochastic regime.

\end{abstract}

\textit{pacs~~{51.60.+a, 78.20.-e, 52.38.Dx}}


\section{Collisionless absorption in regular regime.}
One of the novel problems of laser-matter interaction is the problem of energy absorption in nanometer targets subjected to an ultrashort (up to
few picosecond) intence ($10^{13}-10^{17}\mbox{W/cm}^2$) laser fields. During such interaction hot (up to several keV energy) classical plasma
bounded in nanoscale volume is produced, which has a life time of about hundreds of femtoseconds. This is dense plasma with the electron density
of $10^{23}\mbox{cm}^{-3}$ and more. Such systems are used to be called nanoplasma since first experiments of intense short laser interaction
with three-dimensional nanobodies (atomic Van-der-Vaals clusters) were hold in 1996 \cite{ditm96}.

Nanobodies are known to absorb much more compared to traditional targets like gas or even bulk. The great amount of energy contained in tiny
volume results in breakdown, birth of energetic particles and high harmonics generation \cite{jasapara, tom}. Different mechanisms of absorption
were suggested to explain such phenomena. They are inner ionization, inverse bremsstrahlung effect, vacuum heating, collisionless heating and
some others a bit more sophisticated.

As far as nanoplasma is a strongly bounded system with the width much less than laser wave length, the most interesting mechanism of energy
absorption in it is collisionless heating in self-consistent potential. It was considered recently in one-dimensional systems corresponded to
irradiated films and more deeply in three-dimesional systems which correspond to nanoclusters; the important role of it in the absorption
process was evidently shown (for 1D situation see \cite{iop}).

The problem of collisionless energy absorption in thin films irradiated by intence short laser pulse was considered in \cite{iop} in the frames
of the following model. First, the incompressible liquid approximation for the electronic cloud was used\footnote{For more details about the
applicability of this assumption see \cite{iop}.} and both the self-consistent potential and distribution function were taken as if they are
known function slowly changing on times of the laser pulse duration. This means that the self-consisted system of Boltsman and Laplace equations
was supposed to be solved elsewhere. Then, in \cite{iop} dipole aproximation\footnote{System has linear dimensions much less than the laser wave
length, $\lambda=800$nm for typical Ti:Sa laser.} and the perturbation theory on the small parameter of the dimensionless oscillation amplitude
of the electron cloud\footnote{For the reasonable parameters of the system, such as electronic density $10^{23}\mbox{cm}^{-3}$, laser intensity
$5\cdot10^{16}\mbox{W/sm}^{2}$, ocsillation amplitude has the order of $1$nm, and the typical width of the film is $100$nm. See \cite{iop} for
details.} was used. Here we use the same model. The calculation of the rate of collisionless absorption presented in \cite{iop} was based on
quantum-mechanical approach in quasiclassical limit. Note, that the system considered is classical, and the final result in \cite{iop} does not
contain Plank constant. Althogh this method is non-contradictory, it hides some classical features of the system. The present paper has the aim
to make the same calculation for this system in the frames of classical mechanics and to learn more about it possible behaviour.

Let the particle is bounded in self-consistent potential $U(z)$, with the energy distribution function $\F(\epsilon)$. Mean absorbed energy is
defined as: \beq \overline{q}=\int \rmd\epsilon \F(\epsilon)q(\epsilon), \label{1D_q_init_classical}\eeq where $q(\epsilon)$ -- is the work of
the field over the particle with energy $\epsilon$ in the time unit, averaged with the initial condition for this particle. Due to the shielding
effect external \beq \vecE(t)=\vecE_0\cos(\omega t). \label{laser} \eeq and internal $\mathcal{E}$ laser fields differs, and in the model of
incompressible fluid used in \cite{iop} there is a relation\beq {\mathcal{E}}_z={E_{0z}\over{\sqrt{(1-\omega_{\rm p}^2/\omega^2)^2+
4\Gamma^2/\omega^2}}}, \label{internal-external_laser}\eeq   where $\omega_{\rm p}$ is plasma frequency, $\Gamma$ is the damphing constant, and
we direct electric field along $z$-axis.

One-particle Hamilton function of the system considered has the form: \beq \Ham(p,z,t)=\frac{p^2}{2m_{\rm e}}+U(z)-e{ \mathcal{E}_z}z\cos(\omega
t+\alpha) \equiv \Ham_0+V_{\rm int}. \label{H_init}\eeq  It is convenient to come to the action-angle variables $(\I,\varTheta)$, defined for
the non-perturbed system in a standart way \cite{LL1}: \beq \I=\frac{1}{2\pi}\oint\sqrt{2m_\rme(\epsilon-U(z))}\rmd
z,~~~\varTheta=\frac{\partial \mathcal{S}_{\tiny\texttt{g}}(z,\I)} {\partial\I}, \label{H_peremennye_deistvie-ugol}\eeq where \beq
\mathcal{S}_{\tiny\texttt{g}}(z,\I)=\int\sqrt{2m_\rme (\epsilon(\I)-U(z))} \rmd z \label{H_proizv_funkciya}\eeq is a generating function. Using
the expansion of the coordinate $z(t)$ in Fourier series \begin{eqnarray} z(\I,\varTheta)=\sum_n z_n(\I)\cos{n\varTheta},~~~\nonumber\\ \mbox{so
that~}z(\I,\Theta)\cos\omega t=\sum_n z_n(\I)[\cos(n\Theta+\omega t)+\cos(n\Theta-\omega t)] \label{H_razlozheniye_z} \end{eqnarray} we omit
highly oscillating terms. Then Hamilton function looks like \beq \Ham(\I,\varTheta)=\Ham_0(\I)-\frac{e{ \mathcal{E}}_z}{2}\sum_sz_{s}(\I)
\cos((2s+1)\varTheta-\omega t),~~z_n\equiv z_s. \label{H_all} \eeq

Slow change of the one-particle distribution function on large time scale is a diffusion in action (energy) space (see, for example,
\cite{lihtenberg}), and may be defined by the Fokker-Plank-Kolmogorov equation \beq \frac{\partial\F(\epsilon,t)}{\partial
t}=\frac{1}{2}\frac{\partial} {\partial\epsilon}\D(\epsilon)\frac{\partial\F(\epsilon,t)}{\partial\epsilon}.\label{fpk-equation}\eeq Here we
first introduce the diffusion coefficient \beq \D=\frac{\langle(\delta\I)^2\rangle_{\texttt{T}}}{{\texttt{T}}},\label{diffusion_coefficient}\eeq
which is defined on large (ideally infinite) observation time ${\T}$, and do not depend on it. $\delta \I$ is the addition to unperturbed action
under the influence of the perturbation in (\ref{H_init}). To define it we should write down and solve motion equations:$$ \dot \I=-\frac{e{
\mathcal{E}}_z}{2}\sum_s(2s+1)z_{s}(\I)\sin\varPsi \nonumber
$$
\beq \dot\varTheta=\Omega(\I)-\frac{e{ \mathcal{E}}_z}{2}\sum_s\Omega(\I)z^{\prime}_{s}(\I) \cos\varPsi,~~~~~\varPsi=(2s+1)\varTheta-\omega t.
\label{H_ur_dvizheniya_sum} \eeq Then \beq \delta \I_1=\frac{e{ \mathcal{E}}_z}{2}\sum_s(2s+1)z_{s}(\I_0)\cdot \frac{\cos\beta_s
t-1}{\beta_s},~~~~~ \beta_s=(2s+1)\Omega(\epsilon_0)-\omega, \label{resh_ur_dv_sum}\eeq and finally for the diffusion coefficient
(\ref{diffusion_coefficient}) we get \beq\D=\frac{e^2\mathcal{E}^2}{4}\sum_s\sum_r(2s+1)(2r+1)z_s(\I)z_r(\I)\frac{\langle(1-\cos\beta_s
\T)(1-\cos\beta_r \T)\rangle_{\T}}{\T\beta_s\beta_r}. \label{D_calculation_init}\eeq To get the energy gain in regular regime without
stochasticisy in such description we should take the formal limit $T\to\infty$, because nothing can change the particle trajectory in
collisionless system. During the averaging procedure delta-function and delta-symbol $\delta_{rs}$ appears, and finaly diffusion coefficient in
regular regime reads as \beq
\D(\I)=\frac{\e^2\mathcal{E}^2}{4}\sum_s(2s+1)^2z_s^2(\I)\pi\delta(\beta_s).\label{diffusion_coefficient_regular}\eeq To obtain energy gain we
integrate FPK-equation (\ref{fpk-equation}) by parts and come from action to energy we get:
\begin{eqnarray}
  q(\epsilon)=\frac{\partial\langle\epsilon\rangle}{\partial t}=\left\langle\frac{\partial\I}{\partial t}\frac{\partial\epsilon}{\partial\I}\right\rangle=
  \frac{\partial}{\partial t}\int\I~\rmd\I~\Omega(\I)\F(\I),~~~~\mbox{and}~~~~~~~~~~~~~~~~~~~~~~~~~~~~~~~\nonumber\\
  \frac{\partial\langle\epsilon\rangle}{\partial
  t}=\frac{1}{2}\int\I\Omega(\I)\frac{\partial}{\partial\I}\D(\I)\frac{\partial\F(\I)}{\partial\I}\rmd\I=-
  \frac{1}{2}\int\rmd\I~\D(I) \Omega(I) \frac{\partial\F(\I)}{\partial\I}=\nonumber\\=-\frac{\omega a\mathcal{E}_z^2}{8}m_\rme\omega_{\rm p}^2\sum_s {z_s^2(\I_s)\over
\vert\Omega^{\prime}_{\epsilon}(\epsilon_s)\vert} \frac{\partial\F(\I(\epsilon_s))}{\partial\epsilon}. \label{energy_gain_regular}\end{eqnarray}
This result repeats the result from \cite{iop}, where it was obtained on the quasiclassical language. The sum over $s$ here means that only the
particle at resonant levels \beq \omega=(2s+1)\Omega(\epsilon)\label{1D_condition}\eeq can absorb energy.

\section{Chirikov criterion of the stochasticity.}

The technic presented in the previous section allows to describe classical motion of the particles in nanoplasma in the region of parameters,
where quasiclassical description used in \cite{iop} fails. Supposing that the particle has the initial energy close to resonance energy defined
by the condition (\ref{1D_condition}), averaging over time (\ref{H_init}) we obtain so-called resonant Hamilton function \beq
\Ham(\I,\varTheta)=\Ham_0(\I)-\frac{e{ \mathcal{E}}_z}{2}z_{s}(\I) \cos((2s+1)\varTheta-\omega t),~~z_n\equiv z_s. \label{H_resonant} \eeq
According to \cite{chirikov} let us carry out one more canonical transformation with the help of generating function \beq
\mathcal{G}(\I,\varPsi,t)=-(\I-\I_{s})\frac{\varPsi+\omega t}{2s+1},\label{H_proizv_Funkciya}\eeq with $\I_{s}$ -- resonant action, corresponded
to the resonance energy with the number $s$ from (\ref{1D_condition}). For new $(P,\varPsi)$ and old $(\I,\varTheta)$ variables the following
relations are fulfilled \beq P=-\frac{\partial \mathcal{G}}{\partial\varPsi}=\frac{\I-\I_{s}}{2s+1},
~~~\varTheta=-\frac{\partial\mathcal{G}}{\partial \I}=\frac{\varPsi+\omega t}{2s+1}\label{H_proizv_funk_sootnosheniya}\eeq In new variables
Hamiltonian (\ref{H_resonant}) has the form \beq
\Ham(P,\varPsi)=\frac{1}{2}(2s+1)^2\Omega_\epsilon^\prime(\epsilon_{s})\Omega(\epsilon_{s})P^2-\frac{e\mathcal{E}_z}{2}z_s(\epsilon_{s})\cos\varPsi,
\label{H_nonlinear_osc}\eeq where the decomposition on the small deviation from the resonance action is presented. Hamiltonian function
(\ref{H_nonlinear_osc}) describes the nonlinear mathematical pendulum. Chaotic motion begins when the amplitude in action of such a system is
greater than the distance between neighbour
\begin{figure}[h]
\centerline{\includegraphics[width=10.cm,clip=true]{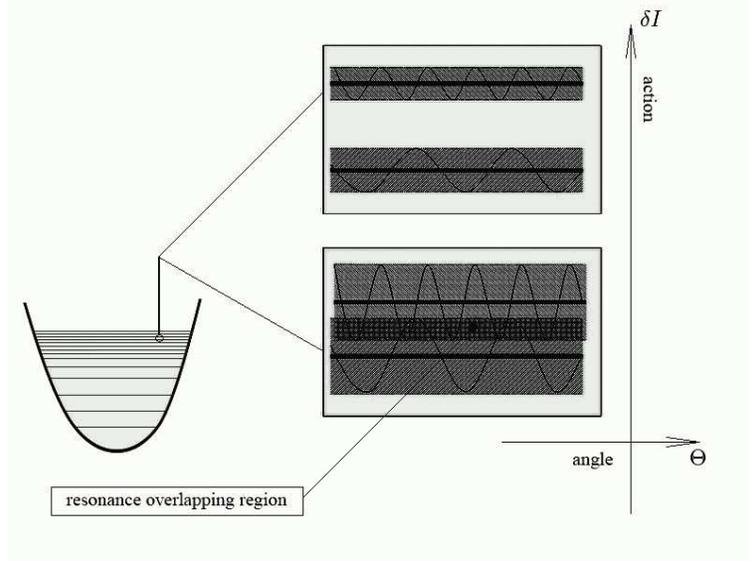}} \caption{The resonances overlapping.}\label{Fig_resonansy}
\end{figure}%
resonances\footnote{The Chirikov criterion of the resonances overlapping.} \cite{chirikov}: \beq \Delta
\I=\sqrt{\frac{2e\mathcal{E}_zz_s(\epsilon_s)}{\left|\Omega_\epsilon^\prime(\epsilon_s)\right|\Omega(\epsilon_s)}}\gtrsim
|\I(\epsilon_s)-\I(\epsilon_{s+1})|. \label{H_kriterij_I}\eeq This condition means that the action oscillation near one resonance come to the
region, occupied by the nearest another one (see Fig.\ref{Fig_resonansy}). The distance between resonances in terms of frequences, according to
condition (\ref{1D_condition}) is \beq |\Omega(\epsilon_s)-\Omega(\epsilon_{s+1})|=\left|\frac{\omega}{2s+1}-
\frac{\omega}{2s+3}\right|\approxeq \frac{\omega}{2s^2}.\label{H_rasstoyanie_mezhdu_resonansami}\eeq Then, with (\ref{H_kriterij_I}), the
criterion of the stochastity appearance, expressed through the field strength inside the system is: \beq
\mathcal{E}_z\gtrsim\frac{1}{|e|}\left|\frac{\omega^2} {8z_s(\epsilon_s)\Omega_\epsilon^\prime(\epsilon_s)\Omega(\epsilon_s)s^4}\right|.
\label{H_kriterij_omega} \eeq This is overestimation: when the inequality is fulfilled, chaos {\textsl{has}} to take place. In reality threshold
is lower \cite{chirikov}. For model potentials from (\ref{H_kriterij_omega}) one can find that in rectangular well of the 100 nm width chaotic
regime begins if the inner field (\ref{internal-external_laser}) is about $10^{15}\mbox{W/cm}^2$.

\section{Collisionless absorption in stochastic regime.}
In the situation when chaotic behaviour is developed, to calculate the heating rate in (\ref{D_calculation_init}) it is nesessary to use for
averaging not infinite time, but such a time $\tau_c$ which define the dynamical memory of the particle about its previous history. The standart
procedure to find it is to define mapping of angle variable \cite{zasl}, which in our system is \beq
\Theta_\nn=\Theta_\n+\frac{2\pi\Omega(\epsilon)}{\omega}+\K\sum_s\chi_s\sin((2s+1)\Theta_\n+\alpha_s), \label{mapping}\eeq where
\begin{multline}
\K=\sum_s\sqrt{\A_s^2+\B_s^2}, ~~~~~~ \chi_s=\sqrt{\A_s^2+\B_s^2}/\sum_s\sqrt{\A_s^2+\B_s^2},\\ \alpha_s=
\frac{(2s+1)\pi\Omega}{\omega}\arccos\left[\frac{\B_s}{\sqrt{\A_s^2+\B_s^2}}\right],\\
\A_s=-\frac{e\mathcal{E}_z}{\omega-(2s+1)\Omega} \left(\left(\frac{\partial
z_s(\I)}{\partial\I}+\frac{z_s(2s+1)}{\omega-(2s+1)\Omega}\frac{\partial\Omega}{\partial\I}\right)\sin((2s+1)\pi\Omega/\omega)\right. + \\
\left. + \frac{z_s(2s+1)\pi\cos((2s+1)\pi\Omega/\omega)}{\omega}\frac{\partial\Omega}{\partial\I}\right)\\ \B_s=
\frac{e\mathcal{E}_z}{\omega-(2s+1)\Omega}\frac{3\pi(2s+1)z_s\sin((2s+1)\pi\Omega/\omega)}{\omega} \frac{\partial\Omega(\I)}{\partial\I}
\label{mapping_descriptions}.\end{multline} For the step time of mapping (\ref{mapping}) the period of external laser field was taken.  For such
mapping the decorellation time $\tau_c$ can be estimated as \cite{zasl} \beq \tau_c\simeq\frac{4\pi}{\Delta s~\omega\ln\K},\label{tau_c}\eeq
where $\Delta s$ is the number of essential items $\chi_s$ in sum (\ref{mapping}). In our situation it has the order of $2s+1$. Finally, the
diffusion coefficient can be obtained from (\ref{D_calculation_init}) with substitution $\T\rightarrow\tau_c$. It reads \beq
\D(\I)=\frac{e^2\mathcal{E}^2}{4}\sum_s\sum_r(2s+1)(2r+1)z_s(\I)z_r(\I)\Delta_{rs}(\tau_c), \label{D_stochastic}\eeq where \beq
\Delta_{rs}(\tau)=\frac{(1-\cos\beta_s \tau)(1-\cos{\beta_r \tau})}{\tau\beta_s\beta_r}.\label{Delta}\eeq Energy gain can be obtained from the
FPK-equation in the same way as we did it earlier for regular regime (\ref{energy_gain_regular}): \beq
\frac{\partial\langle\epsilon\rangle}{\partial t} = \D(\I)=\frac{e^2\mathcal{E}^2}{4}\int \rmd\epsilon \Omega(I(\epsilon))
\frac{\partial\F(\I(\epsilon))}{\partial\epsilon} \sum_{sr}(2s+1)(2r+1)z_s(\I(\epsilon))z_r(\I(\epsilon))
\Delta_{rs}(\epsilon,\tau_c).\label{energy_gain_stochastic}\eeq In this expression all resonance levels take part in the absorption process
simultaneously. Moreover, in such a situation particle with arbitrary energy should gain energy from the external field. Formula
(\ref{energy_gain_stochastic}) is the main result of the present work. It describes the collisionless heating in 1D classical nanoplasma layer
when the field strength is enough for chaotic regime to take place, according to (\ref{H_kriterij_omega}).

Author wishes to thank S.V. Popruzhenko, D.F. Zaretsky, I.Yu. Kostyukov for fruitful discussions. The work was done with the financial support
of RFBR.


\vskip 2cm

\end{document}